\documentstyle[twocolumn,prl,aps]{revtex}

\begin{document}

\draft

\title{No well-defined remnant Fermi surface in
Sr$_2$CuO$_2$Cl$_2$}

\author{S. Haffner\cite{author}, D. M. Brammeier, C. G. Olson, L. L. Miller,
and D. W. Lynch}
\address{Department of Physics and Astronomy and Ames Laboratory,\\
U. S. Department of Energy, Iowa State University,\\
Ames, Iowa 50011}

\date{\today}

\maketitle

\begin{abstract}
In angle-resolved photoelectron spectra of the antiferromagnetic
insulators Ca$_2$CuO$_2$Cl$_2$ and Sr$_2$CuO$_2$Cl$_2$ a sharp
drop of the spectral intensity of the lowest-lying band is
observed along a line in {\bf k} space equivalent to the Fermi
surface of the optimally doped high-temperature superconductors.
This was interpreted as a signature of the existence of a remnant
Fermi surface in the insulating phase of the high-temperature
superconductors. In this paper it is shown that the drop of the
spectral intensity is not related to the spectral function but is
a consequence of the electron-photon matrix element.
\end{abstract}

\vspace{1cm}


It is of interest to understand the changes the normal-state
electronic structure of the high-temperature superconductors
(HTSCs) undergoes when doped from an antiferromagnetic insulator
to a metal. Angle-resolved photoelectron spectroscopy (ARPES) has
played an important role in the study of the electronic structure
of the HTSCs in the different regions of the doping phase
diagram.  Recently ARPES
experiments using Sr$_2$CuO$_2$Cl$_2$ and Ca$_2$CuO$_2$Cl$_2$
single crystals yielded information on the electronic structure
of the HTSCs in the limit of very low
hole doping.\cite{Wells95,LaRosa96,Kim98,Ronning98}
\par
Sr$_2$CuO$_2$Cl$_2$ and Ca$_2$CuO$_2$Cl$_2$ are closely related
to the undoped parent compounds of the HTSCs, as they are also
antiferromagnetic insulators with N\'{e}el temperatures of 255
and 247 K, respectively.\cite{Greven94,Vaknin97} The CuO$_2$
planes in Sr$_2$CuO$_2$Cl$_2$ and Ca$_2$CuO$_2$Cl$_2$ are undoped
(half-filled); therefore the spectral intensity in an
electron energy-distribution curve (EDC) related to the CuO$_2$
plane gives information about the dynamics of a single hole (the
hole created by photoionization) in a CuO$_2$ plane. Of primary
interest are the lowest-lying states in an ARPES spectrum, the
so-called first electron-removal states. 
In Sr$_2$CuO$_2$Cl$_2$ and
Ca$_2$CuO$_2$Cl$_2$ these show a well-developed low-binding-energy
peak only for {\bf k} vectors in the vicinity of ($\pi$/2,$\pi$/2)
where the peak has its minimum binding
energy.\cite{Wells95,LaRosa96,Kim98,Ronning98,Annotation1} The
low-binding-energy peak is followed by additional spectral weight
at higher binding energies. Outside this {\bf k} space region the
first feature is a rather broad structure with its maximum located
at higher binding energy.
\par
In the published Sr$_2$CuO$_2$Cl$_2$ and Ca$_2$CuO$_2$Cl$_2$
ARPES data the spectral intensity of the first electron-removal
states shows a sudden drop along a line in {\bf k} space which
closely resembles the Fermi surfaces found in band-structure
calculations or ARPES data of optimally or overdoped
HTSCs.\cite{Ronning98} Insulators such as Sr$_2$CuO$_2$Cl$_2$,
Ca$_2$CuO$_2$Cl$_2$ and the undoped parent compounds of the
HTSCs do not have partially filled bands and hence no Fermi
surface, but it was argued that there exists a remnant of the
Fermi surface of their respective metallic phases in the sense
that when the first electron-removal states of
Sr$_2$CuO$_2$Cl$_2$ or Ca$_2$CuO$_2$Cl$_2$ cross this remnant
Fermi surface, their spectral intensity suddenly drops, in analogy
to the behavior of the ARPES spectral intensity related to the
lowest-lying band of a metal when it crosses the Fermi surface.
Comparing the binding energies of the first electron-removal
states of the copper oxychlorides for {\bf k} vectors on the
remnant Fermi surface one can formally define a gap, zero
for {\bf k} on the remnant Fermi surface where the
low-binding-energy peak has minimum binding energy, and
maximum for {\bf k} on the remnant Fermi surface
where the low-binding-energy peak has maximum binding energy. It
was noted that the gap defined this way shows an analogous {\bf k}
dependence as the superconducting gap of the HTSCs, i.e. a
$d$-wave-like dispersion.\cite{Ronning98} This $d$-wave-like
behavior of the first electron-removal states of the copper
oxychlorides would naturally explain the {\bf k} dependence
of the normal-state gap encountered in the underdoped HTSCs:
upon hole-doping the chemical potential drops to the maximum of
the $d$-wave-like dispersion (the node of the "gap") of the
insulator. In the underdoped regime the Fermi level only touches
states near the node of the $d$-wave-like gap of the insulator,
forming small segments of the Fermi surface, while the portions of
the remnant Fermi surface around ($\pi$,0) remain gapped.
\par
However, in ARPES, the spectral intensity of
an EDC is directly proportional to the electron-photon matrix-element weighted
spectral function \cite{Randeria95} and not the spectral function itself.
Calculations suggest that the relationship between ARPES intensities and the
underlying electronic structure can be complicated due to matrix
element effects and that caution should be exercised in
interpreting detailed features of the ARPES intensities in terms
of the spectral function.\cite{Bansil98} In a recent paper, using
ARPES of Sr$_2$CuO$_2$Cl$_2$ as an
example, it was demonstrated experimentally that
for layered cuprates the electron-photon matrix element can have a
significant impact on both the relative spectral intensity and
the shape of a feature in an ARPES spectrum.\cite{Haffner00} In
this paper ARPES measurements on a Sr$_2$CuO$_2$Cl$_2$ single
crystal along the line $\Gamma$ to ($\pi$,$\pi$) in {\bf k} space,
using different photon energies in the range from 20 to 24 eV, are
presented. It will be shown that the sudden drop in spectral
intensity observed in the ARPES spectra of Sr$_2$CuO$_2$Cl$_2$
along a line in {\bf k} space similar to the Fermi surface of
HTSCs is not due to a similar drop in the spectral function due
to a remnant Fermi surface but is caused by the electron-photon
matrix element i.e. is not a feature of the electronic structure
of these compounds but an artifact of the photoexcitation.
\par
The ARPES were recorded on the
 Ames Laboratory/Montana
State University ERG/SEYA beamline at the Synchrotron Radiation
Center, using a 50 mm radius hemispherical analyzer with a
2$^{\circ}$ full angular acceptance angle, corresponding to a
{\bf k} resolution of 0.034 \AA$^{-1}$ and 0.038 \AA$^{-1}$ (6 \%
and 6.8
\% of the distance between $\Gamma$ and ($\pi$,$\pi$)) for the
first electron-removal states and 20 eV and 24 eV photon energy,
respectively. The total energy resolution was 105 meV. The angle
of incidence of the photons was $\sim$43$^{\circ}$ with respect
to the sample surface normal with the sample normal pointing down
by $\sim$8$^{\circ}$. The Sr$_2$CuO$_2$Cl$_2$ single crystal was
grown as described elsewhere \cite{Miller90}. The sample was
oriented {\it ex situ} by Laue backscattering and mounted with
the Cu-O bonds in a horizontal/vertical plane. The sample
was cleaved (cleavage plane parallel to CuO$_2$ planes) in the
experimental chamber in a vacuum better than 4$\times$10$^{-11}$
Torr and sample alignment was confirmed {\it in situ} by
using the symmetry of the dispersion of spectral features at
high-symmetry points and the appearance of acute peaks associated
with purely O 2$p$-derived states in the main valence band at
$\sim$2.5 and $\sim$4 eV at ($\pi$,$\pi$) and ($\pi$,0),
respectively.\cite{Pothuizen97} All EDCs were recorded at room
temperature and were normalized to the photon flux. Although
there is no long-range antiferromagnetic order at room
temperature in Sr$_2$CuO$_2$Cl$_2$, the antiferromagnetic
correlation length is still two orders of magnitude larger than
the Cu-O distance \cite{Greven94}. Therefore photoemission, as a
fast and local probe, still sees the effect of antiferromagnetic
order, even 50 K above the N\'{e}el temperature \cite{Wells95}.
The Fermi edge of a Pt foil in electrical contact with the
sample was used as binding-energy reference. There were no
indications of charging effects because repeating an EDC
 after the beam had decayed to less than half the
current when the first EDC was taken gave the same EDC, when
normalized to the incident flux. All spectra shown were
recorded within a period of five days after the cleave of the
sample and in one experimental run, i.e., using the same sample
cleave. We observed no signs of sample degradation during this
span of time.
\par
Figure 1 presents the first electron-removal states of
Sr$_2$CuO$_2$Cl$_2$ for {\bf k} along $\Gamma$ -
($\pi$,$\pi$) for photon energies from 20 to 24 eV. For all
photon energies a dispersing low-binding-energy peak is observed
which has its minimum binding energy at ($\pi$/2,$\pi$/2). There
is a significant change in the spectral intensity of the
low-binding-energy peak as a function of {\bf k} for different
photon energies. For 24 eV, the low-binding-energy
peak has non-negligible spectral intensity for {\bf k} 
from 38 to 51 \% of the distance between $\Gamma$ and
($\pi$,$\pi$). Upon lowering the photon energy, the center of the
{\bf k } span in which this peak exists along
the line $\Gamma$ to ($\pi$,$\pi$) shifts to bigger $|${\bf k}$|$.
For example, for 22 eV photons the low-binding-energy peak can be
observed for {\bf k } from 39 to 58 \% of the distance
between $\Gamma$ and ($\pi$,$\pi$), while for 20 eV photon energy,
it has significant spectral intensity from 49 to 60
\% of the distance between $\Gamma$ and ($\pi$,$\pi$). This trend
is exemplified in Fig. 2, which shows the integrated spectral
intensity of the low-binding-energy peak associated with
the momentum density $n({\bf k})$. It is evident that the maximum
of $n({\bf k})$ shifts from $\sim$40 to 55 \% of the distance
between $\Gamma$ and ($\pi$,$\pi$) on going from 24 to 20 eV
photon energy. This is a substantial shift, given the fact that
along $\Gamma$ - ($\pi$,$\pi$) in {\bf k} space there is only a
low-binding-energy peak from $\sim$33 to $\sim$60 \% of the
distance between $\Gamma$ and ($\pi$,$\pi$).
\par
Note that the observed differences between the series of EDCs
shown in Fig. 1 can only be related to the electron-photon matrix
element. Sample variability can be excluded as the spectra were
recorded using the same sample and cleave. Furthermore,
our data are compatible with previously published
Sr$_2$CuO$_2$Cl$_2$ and Ca$_2$CuO$_2$Cl$_2$ ARPES results (if
recorded at the same photon energy) which eliminates a bad sample
cleave or misalignment of the crystal as possible reasons. One
could argue that, dependent on photon energy, different initial
states are observed, but this is not consistent with the continous
shift of spectral intensity of the low-binding-energy peak to
bigger {\bf k} vectors evident in Fig. 1 when the photon energy
is decreased. Moreover, up to now most authors ascribe the
low-binding-energy peak evident in the first electron-removal
states of Sr$_2$CuO$_2$Cl$_2$ or Ca$_2$CuO$_2$Cl$_2$ to a
Zhang-Rice singlet (ZRS) \cite{Zhang88,DagottoReview}, i.e. one
initial state. For an initial state with a given {\bf k}
component parallel to the CuO$_2$ planes the momentum component
perpendicular to the CuO$_2$ planes is dependent on the photon
energy but a ZRS is localized in a CuO$_2$ plane; therefore no
dependence of the dispersion of the low-binding-energy peak on
the momentum component perpendicular to the CuO$_2$ plane is
expected. Note also that the in-plane nature of the counterpart of
the first electron-removal states in the unoccupied part of the
electronic structure (which is derived from the same orbitals as
the first electron-removal states) has been explicitly shown by
x-ray absorption spectroscopy.\cite{Haffner98} The spectral
function itself does not depend on the photon energy used to
excite the first electron-removal states. The $\Gamma$ -
($\pi$,$\pi$) series of spectra shown in Fig. 1 are equivalent in
the sense that they show the lowest-lying excitations associated
with the motion of a hole in an antiferromagnetically ordered
CuO$_2$ plane for {\bf k} along the $\Gamma$-
($\pi$,$\pi$) line in the first Bz, i.e. the underlying
spectral function is the same. The electron-photon matrix element
$\langle$ i $|$ {\bf p}$\cdot${\bf A} $|$ f $\rangle$ ({\bf p}
and {\bf A} are the photoelectron momentum and the vector
potential, respectively), on the other hand, is affected by the
photon energy as different final states are reached upon changing
the photon energy.  It therefore is the factor which is responsible
for the observed differences between the series of EDC's shown in
Fig. 1.
\par
From the preceeding discussion we can
conclude that the spectra shown in Fig. 1 truly represent the
low-binding energy ARPES response of Sr$_2$CuO$_2$Cl$_2$ for 20 to
24 eV photon energy and {\bf k} from $\Gamma$ to
($\pi$,$\pi$) and it has become clear that the spectra are
significantly influenced by the electron-photon matrix element. 
These data show that the lowest-lying band does not cross a sharp
remnant of a Fermi surface. Firstly, note that in this kind of
scenario the lowest-lying band would disperse to lower binding
energies on going from $\Gamma$ to ($\pi$/2,$\pi$/2) and cross the
remnant Fermi surface before it disperses back to higher binding
energies again i.e. for {\bf k} before ($\pi$/2,$\pi$/2).
This implies that in Fig. 1 the peak associated with the
lowest-lying band should lose most of its spectral intensity
before ($\pi$/2,$\pi$/2). This is the case for 24 and 23 but not
for 20 and 21 eV photon energy where this
 peak has most of its spectral intensity for
{\bf k} between ($\pi$/2,$\pi$/2) and ($\pi$,$\pi$).
According to Ref.\ \onlinecite{Ronning98}, a way to locate the
crossing of the Fermi surface (or remnant Fermi surface) is to
determine the {\bf k} for which n({\bf k}) of the
low-binding-energy peak as a function of {\bf k} has its steepest
descent. Applying this method on our Sr$_2$CuO$_2$Cl$_2$ ARPES
data, it is clear that due to the influence of the electron-photon
matrix element we do not get a result which is robust for 20 to 24
eV photon energy as the {\bf k} where there is the steepest
descent in n({\bf k})  from the low-binding-energy peak along
$\Gamma$ to ($\pi$,$\pi$) (Fig. 2) varies from $\sim$50 to
$\sim$63 \% of the distance between $\Gamma$ to ($\pi$,$\pi$) for
24 to 20 eV photon energy, which, as already noted, is a
substantial shift compared with the {\bf k} space range where
there exists a low-binding-energy peak in the first
electron-removal states of Sr$_2$CuO$_2$Cl$_2$.
\par
The fact that there is a distinct low-binding-energy peak for
{\bf k} on a line from ($\pi$/2,$\pi$/2) to ($\pi$,$\pi$)
and that the steepest descent of the integrated spectral
intensity of the low-binding-energy peak along $\Gamma$ to
($\pi$,$\pi$) is not found at the same {\bf k} for
different photon energies represents evidence against the
existence of a well-defined remnant Fermi surface for
Sr$_2$CuO$_2$Cl$_2$. Nonetheless, the existence of a remnant Fermi
surface for Sr$_2$CuO$_2$Cl$_2$ cannot be ruled out unless it is
also proven that the intensity of the low-binding-energy peak for
{\bf k} along $\Gamma$ - ($\pi$/2,$\pi$/2) is not much bigger
than for {\bf k} along ($\pi$/2,$\pi$/2) - ($\pi$,$\pi$).
Note that this possibility still exists, as the low-binding-energy
peak is best developed for {\bf k} along $\Gamma$ -
($\pi$/2,$\pi$/2) for a different photon energy (23 eV) than for
{\bf k} along ($\pi$/2,$\pi$/2) to ($\pi$,$\pi$) (21 eV).
This means that it still could be possible that the intensity of
the low-binding-energy peak in the spectral function is quite
different for {\bf k} along $\Gamma$ - ($\pi$/2,$\pi$/2)
and along ($\pi$/2,$\pi$/2) - ($\pi$,$\pi$). It is clear that a
much bigger intensity of the low-binding-energy peak for {\bf k}
along $\Gamma$ - ($\pi$/2,$\pi$/2) than for {\bf k}
along ($\pi$/2,$\pi$/2) - ($\pi$,$\pi$) would be
compatible with a sharp decrease of the integrated intensity of
this peak on going from $\Gamma$ to
($\pi$,$\pi$) and a remnant Fermi surface. In Fig. 3 the
first-electron removal states of Sr$_2$CuO$_2$Cl$_2$ are shown
for {\bf k} vectors at 44 and 56 \% of the distance between
$\Gamma$ and ($\pi$,$\pi$) [i.e. equidistant to
($\pi$/2,$\pi$/2)] for photon energies from 20 to 24 eV. If there
exists a remnant Fermi surface in Sr$_2$CuO$_2$Cl$_2$ it is
expected that the spectral intensity of the
peak at 56 \% of the distance between $\Gamma$ and ($\pi$,$\pi$)
and for the photon energy where it is best developed (21 eV) is
much smaller than the maximum peak intensity one can get at 44 \%
of the distance between $\Gamma$ and ($\pi$,$\pi$) (at 23
eV photon energy) which evidently is not the case. In conjunction
with the results presented in the preceeding paragraph this is
clear evidence that in the case of Sr$_2$CuO$_2$Cl$_2$ the
lowest-lying band does not cross a remnant Fermi surface on going
from $\Gamma$ to ($\pi$,$\pi$).
\par
In conclusion the spectral function associated with the motion of
a hole in an antiferromagnetically ordered background has been
probed by ARPES of the
antiferromagnetic insulator Sr$_2$CuO$_2$Cl$_2$. Along
the line $\Gamma$ to ($\pi$,$\pi$) in {\bf k} space we have
observed that the strong drop of the spectral intensity of the
lowest-lying states reported in the literature is not related to a
crossing of the lowest-lying band of a remnant Fermi surface but
is due to the electron-photon matrix element.

\par
The Ames Laboratory is operated by Iowa State University for the
U.S. DOE under Contract No. W-7405-ENG-82. This work is based upon
research conducted at the Synchrotron Radiation Center, University of
Wisconsin, Madison, which is supported under Award No.
DMR-95-31009.

\newpage

\begin{figure}
\caption{The first electron-removal states of Sr$_2$CuO$_2$Cl$_2$
recorded along the $\Gamma$
to ($\pi$,$\pi$) direction of the first BZ using 20 to 24 eV photon energy.
The {\bf k} vectors are given in \% of the distance between $\Gamma$ and
($\pi$,$\pi$).}
\end{figure}

\begin{figure}
\caption{Integrated spectral intensity of the low-binding-energy
peak of the first electron-removal states of Sr$_2$CuO$_2$Cl$_2$
deduced from the spectra shown in Fig. 1. {\bf k} vectors are
given in \% of the distance between $\Gamma$ and ($\pi$,$\pi$).}
\end{figure}

\begin{figure}
\caption{Comparison of the first electron-removal states of
Sr$_2$CuO$_2$Cl$_2$ recorded at 44 (open circles) and 56 \%
(solid circles) of the distance between $\Gamma$ and
($\pi$,$\pi$) for 20 to 24 eV photon energy. The spectra are
offset vertically for clarity but are on-scale otherwise.}
\end{figure}

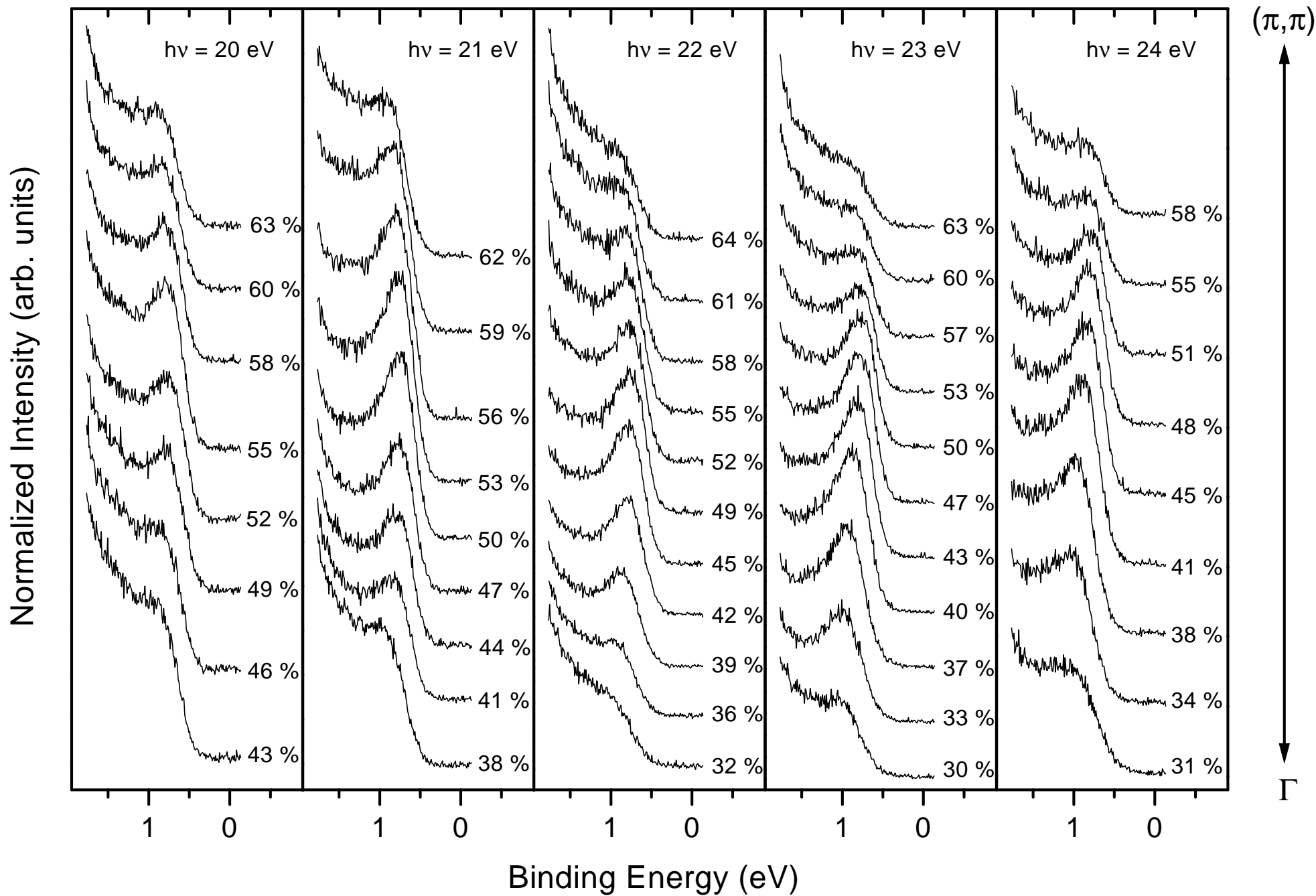
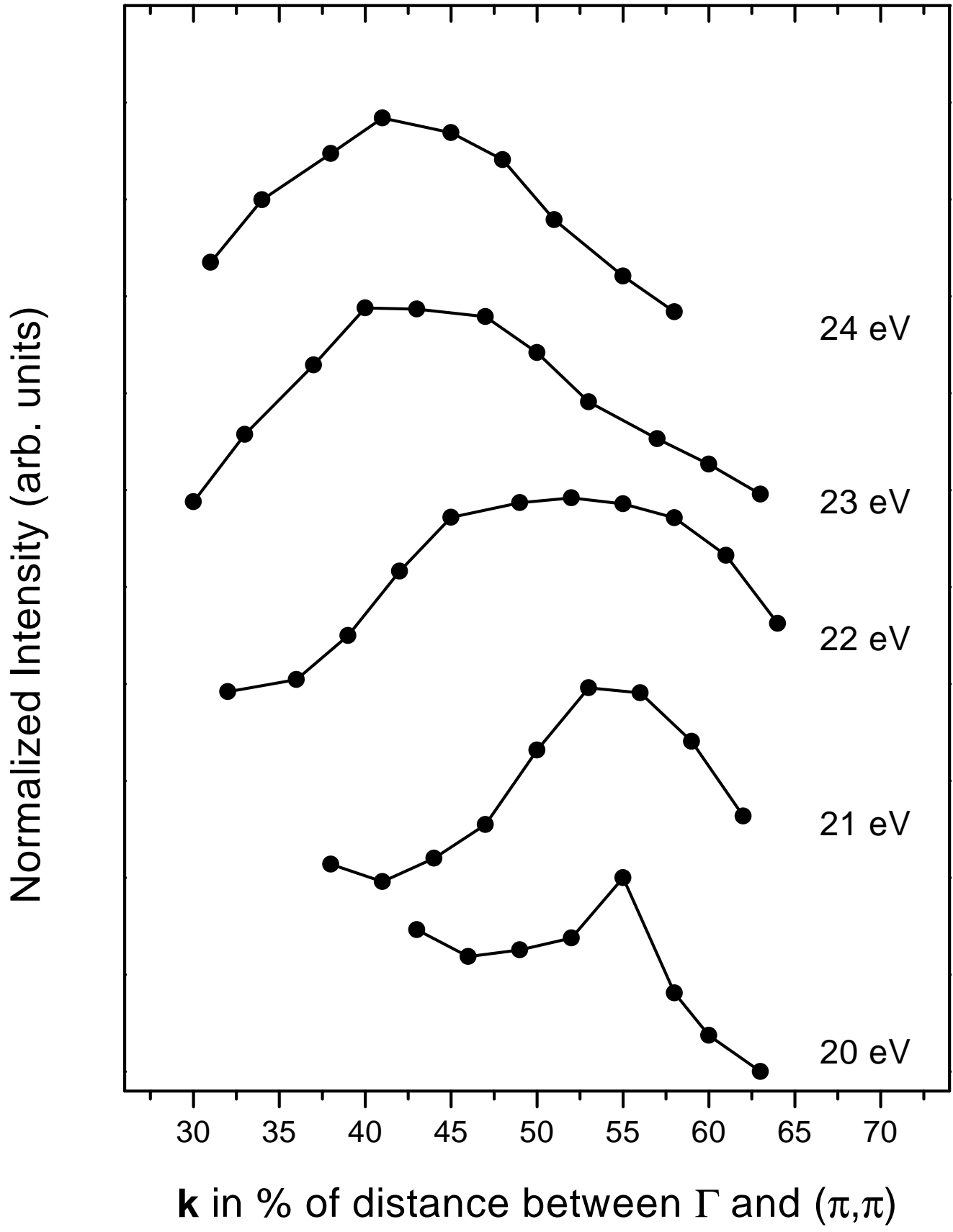
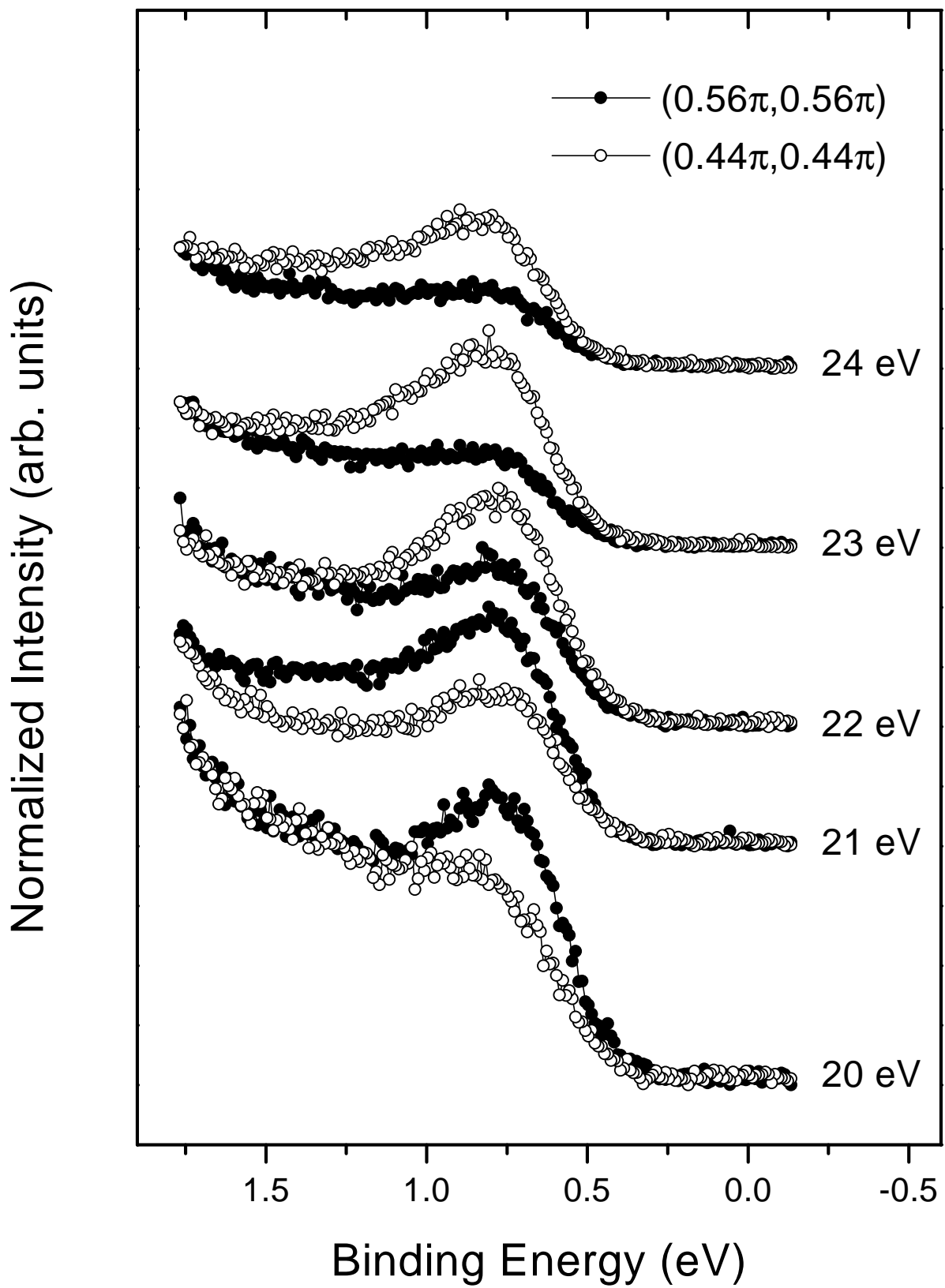
\end{document}